# Kinetics-Limited Two-Step Growth of van der Waals Puckered Honeycomb Sb Monolayer

Zhi-Qiang Shi [1†], Huiping Li [2,3†], Qian-Qian Yuan [1], Cheng-Long Xue [1], Yong-Jie Xu [1], Yang-Yang Lv [4], Zhen-Yu Jia [1], Yanbin Chen [1,5], Wenguang Zhu [2,3]*, Shao-Chun Li [1,5,6]*

*1 National Laboratory of Solid State Microstructures, School of Physics, Nanjing University, Nanjing 210093, China*

*2 International Center for Quantum Design of Functional Materials (ICQD), Hefei National Laboratory for Physical Sciences at the Microscale, and Synergetic Innovation Center of Quantum Information and Quantum Physics, University of Science and Technology of China, Hefei, Anhui 230026, China*

*3 Key Laboratory of Strongly-Coupled Quantum Matter Physics Chinese Academy of Sciences, School of Physical Sciences, University of Science and Technology of China, Hefei, Anhui 230026, China*

*4 National Laboratory of Solid State Microstructures, Department of Materials Science and Engineering, Nanjing University, Nanjing 210093, China*

*5 Collaborative Innovation Center of Advanced Microstructures, Nanjing University, Nanjing 210093, China*

*6 Jiangsu Provincial Key Laboratory for Nanotechnology, Nanjing University, Nanjing 210093, China.*

† These authors contributed equally to this work.

\* Corresponding Authors: scli@nju.edu.cn; wgzhu@ustc.edu.cn.

**ABSTRACT:** Puckered honeycomb Sb monolayer, the structural analog of black phosphorene, has been recently successfully grown by means of molecular beam epitaxy. However, little is known to date about the growth mechanism for such puckered honeycomb monolayer. In this study, by using scanning tunneling microscopy in combination with first-principles density functional theory calculations, we unveil that the puckered honeycomb Sb monolayer takes a kinetics-limited two-step growth mode. As the coverage of Sb increases, the Sb atoms firstly form the distorted hexagonal lattice as the half layer, and then the distorted hexagonal half-layer transforms into the puckered honeycomb lattice as the full layer. These results provide the atomic-scale insight in understanding the growth mechanism of puckered honeycomb monolayer, and can be instructive to the direct growth of other monolayers with the same structure.

The puckered honeycomb lattice emerges as a particular member in the family of two dimensional or layered materials. The well-known example that hosts the puckered honeycomb structure is black phosphorene, which has found extensive interests since its rediscovery in 2014, owing to its exotic properties, such as high carrier mobility, tunable band gap, and anisotropic transport properties, *etc*.[1-14] The bulk black phosphorous can be fabricated through either high pressure or catalyst assistance.[15, 16] However, the single layer of black phosphorous, namely black phosphorene, is only achievable *via* top-down methods of mechanical or chemical exfoliation.[15, 16] To directly grow the black phosphorene is rather difficult, even though the growth of its allotrope, blue phosphorene, has been experimentally and theoretically studied.[17-19] On the other hand, the chemical instability of black phosphorene hinders the further exploration toward the application. An alternative effort is to search for the structural analog of black phosphorene in group V elements, *e.g.*, As, Sb and Bi.[20-27]

The bulk antimony crystallizes in a rhombohedral structure (Space group: $R\bar{3}m$) which is different from the black phosphorous. The corresponding single layer, namely *β*-antimonene, is in the buckled hexagonal lattice and has been well studied.[28-38] Until recently, the large-scale and high-quality puckered honeycomb Sb monolayer, namely *α*-antimonene, has been successfully grown on $T_d$-WTe$_2$ substrate by using molecular beam epitaxy (MBE) technique.[39] Further characterization indicates that the *α*-Sb/WTe$_2$ is very stable upon exposure to air, and exhibits the linear electron band near Fermi energy.[39] To date, a comprehensive understanding for the growth of such puckered honeycomb monolayer is still missing, which should be vital for driving the discovery of more practically useful monolayers with analogous structure.

In this study, we took the MBE-grown single layer *α*-antimonene as an example, to explore its growth mechanism. Through tuning down the growth kinetics, we were able to characterize an intermediate metastable structure for the growing α-antimonene monolayer by using scanning tunneling microscopy (STM). In combination with first-principles density functional theory (DFT) calculations, the growth mechanism was revealed. The growth of α-antimonene can be separated into two stages. Initially, the adsorbed Sb atoms prefer to form a distorted hexagonal (dH)-like lattice, which acts as the half-layer transient state. The transient dH state is energetically metastable and eventually transforms into the puckered honeycomb α-antimonene.

**RESULTS AND DISCUSSION:**

The puckered honeycomb lattice of α-antimonene is composed of two atomic sublayers in vertical corrugation, as depicted in Figure 1a. There are four Sb atoms in each unit cell. Each Sb atom forms three covalent bonds with three neighboring atoms. The freestanding lattice constants of α-antimonene are 4.36 Å and 4.74 Å along zigzag and armchair direction, respectively.[22] Layered semiconductor SnSe crystal (orthorhombic structure, space group: *Pnma*)[40] was adopted as the substrate. The pristine surface of the cleaved SnSe is composed of the micrometer-scale and atomically-flat terrace, with a semiconducting band gap of ~0.85 eV (see Supporting Information Figure S1). Figure 1b shows the surface morphology of the α-Sb monolayer grown on the SnSe substrate kept at ~400 K. The large-scale α-Sb monolayer islands form at this growth condition. Atomically resolved STM image, the inset to Figure 1b, confirms the well-ordered α-Sb structure without obvious defects. In addition, the moiré pattern can be identified in Figure 1b, which originates from the minimal lattice mismatch between Sb and SnSe.

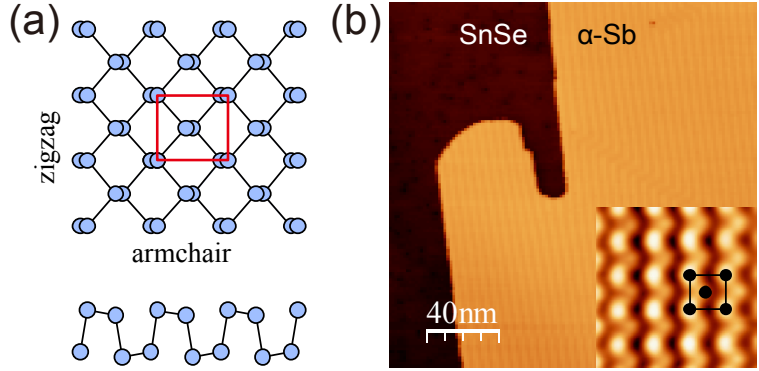

**Figure 1.** Epitaxial growth of single-layer α-antimonene on the SnSe substrate. (a) The crystal structure of the freestanding α-antimonene (upper: top view, lower: side view). The 1 × 1 unit cell is marked by the red rectangle. (b) The STM topographic image (200 × 200 nm$^2$) of the single-layer α-antimonene epitaxially grown on the SnSe substrate. $U$ = +3 V, $I_t$ = 100 pA. Inset: Atomically resolved STM image of the α-antimonene. $U$ = -880 mV, $I_t$ = 5 nA. The 1 × 1 unit cell is marked by the black rectangle.

In order to investigate the growth mechanism in detail, an alternate growth recipe was adopted, *i.e.*, to deposit Sb atoms onto the SnSe substrate kept at a lower temperature of ~350 K. In this way, the diffusion of the surface Sb atoms is limited and the phase transition from the intermediate state to α-antimonene is effectively suppressed. Figure 2a shows the morphology of Sb terrace grown at this condition. In contrast with the smooth Sb layer shown in Figure 1b, the surface becomes corrugated, which is mainly composed of two different regions, *i.e.*, the protrusive and depressive regions. The protrusive region is structurally the same as the full monolayer of α-antimonene. Line-scan profile measurement across the surface, as plotted in Figure 2b, indicates that the step height of protrusive island is ~6.5 Å, consistent with the full monolayer of α-Sb. The depressive region is ~1.6 Å lower than the protrusive region, which is referred to as the half-layer Sb in the following. It is noteworthy that the measured height of the half-layer Sb, ~4.9 Å, is much larger than that of experimentally reported $β$-Sb monolayer.[31] Post annealing of the surface without additional Sb deposition results in the increase of the areal proportion of the full monolayer $α$-Sb, at the expense of the half-layer Sb, as demonstrated in Figure 2c,d. The half-layer regions finally disappear upon continuous annealing (see Supporting Information Figure S2). The STM results taken at various bias voltages in the same area confirm that the surface is stable and no scan-induced structural changes are observed (see Supporting Information Figure S3). It is thus indicated that the half-layer Sb is energetically metastable compared to the full monolayer $α$-Sb, and undergoes a structural transition upon thermally annealing.

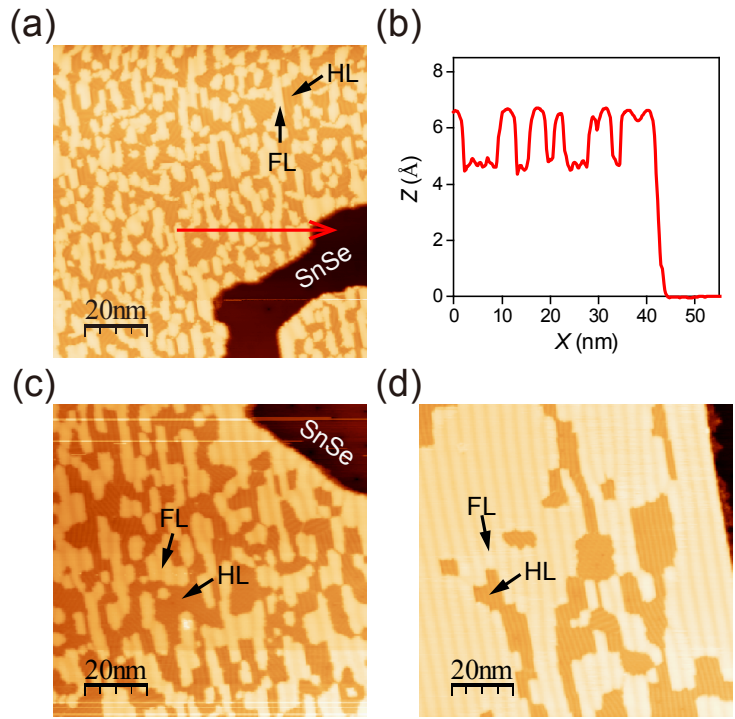

**Figure 2.** Morphology evolution of the Sb film deposited on the SnSe substrate upon post annealing. (a) The as-grown surface (100 × 100 nm$^2$) after ~0.5 MLs Sb are deposited on the SnSe substrate kept at ~350 K. $U = +3$ V, $I_t = 100$ pA. (b) The line-scan profile taken along the red arrowed line in (a). The protrusive and depressed regions, as representatively marked by the arrows in (a, c, d), are assigned to the full layer (FL) of α-antimonene and the half layer (HL) structure of Sb. The step height of the FL and HL Sb regions are ~6.5 Å and ~4.9 Å, respectively. (c, d) The surface (100 × 100 nm$^2$) of the Sb film grown on SnSe after subsequently annealing at ~350 K for 10 min (c) and 30 min (d), respectively. $U = +3$ V, $I_t = 100$ pA.

In the puckered honeycomb lattice, both of the top and bottom atomic layers host the same in-plane orthorhombic symmetry. Figure 3 shows the atomically resolved STM images taken on various half-layer Sb regions. The surface atoms are partially highlighted by yellow dots for better vision. Surprisingly, all the half-layer Sb regions exhibit the distorted hexagonal (dH) symmetry, which is completely different from the full monolayer α-Sb. The moiré patterns of half-layer seem to indicate several different orientations, which can be ascribed to the different stacking order between the dH-Sb and SnSe substrate. Meanwhile, the orientation of the full-layer α-Sb is quite uniform on the substrate. Further structural analysis of the boundary, as shown in Supporting Information Figure S4, indicates that the atomic periodicities of the dH- and α-Sb monolayers are commensurately compatible along the direction of boundary, suggesting that the growth is likely to be seamless under the strain. In addition, there exists slight structural fluctuations among these half-layer regions, which further confirms their metastable nature.

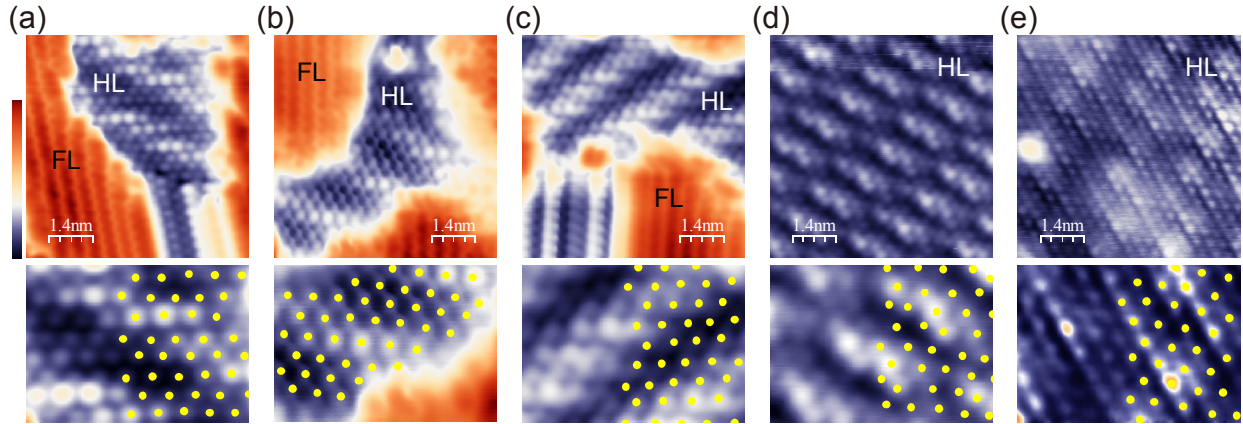

**Figure 3.** Various atomic structures of the half-layer Sb. (a-c) Atomically resolved images taken on the as-grown sample. (d, e) Atomically resolved images taken on the sample after thermally annealing at 350 K for 30 min. Upper panel: large-scale images (7 × 7 nm$^2$). $U$ = -1.3 V; $I_t$ = 1 nA. Lower panel: Zoom-in images (3.5 × 2.5 nm$^2$) extracted from the upper panel. The regions of the half-layer (HL) Sb and full layer (FL) α-Sb are labeled on the images. The yellow dot arrays are partially superimposed on the surface atoms for better vision.

Based on DFT calculations, we identified the dH structure as shown in Supporting Information Figure S5. The calculated density of state of this half-layer dH structure shows a prominent band gap of 0.6 eV. Considering that DFT calculated band gaps at GGA level are normally severely underestimated by roughly half, the calculated band gap is comparable with the STS measured value of 1.0 eV taken on the half-layer region, as displayed in Supporting Information Figure S6. The formation energy calculations also confirm that the dH structure is less stable than the puckered α-Sb monolayer on the SnSe substrate with a formation energy difference of 59 meV/atom. Regarding to the different orientation of morié patterns observed in the half-layer dH-Sb, we performed DFT calculations to evaluate the binding energy of dH-Sb and α-Sb on SnSe substrate with different orientations. The results, as shown in Supporting

Information Figure S7, indicate that the dH-Sb is stable with very similar binding energy in several different orientations. However, the α-Sb dominantly favors the specific orientation and thus is aligned with the SnSe substrate.

We further explored possible kinetic pathways of the structural transformation from the metastable half-layer dH structure to the ground-state full-layer puckered structure. For simplicity, free-standing monolayers were used in the calculations. We first considered the traditional direct process shown in Figure 4, a kinetic pathway similar to that for the transformation from a buckled honeycomb blue P to a puckered black P.[41] During the phase transformation from dH-Sb to *α*-Sb, as illustrated in Supporting Information Figure S8(a), one of every two rows of atoms along the zigzag direction in the upper layer flips to the lower layer, accompanied by one row of the adjacent atoms in the lower layer flipping inversely to the upper layer, without bond breaking. The calculated activation barrier of the process is 525 meV/atom. However, a more feasible kinetic process with a lower activation barrier is revealed *via* a two-step motion as illustrated in Figure 4 and Supporting Information Figure S8(b). In the first step, the dH structure transforms into a metastable intermediate structure through an in-plane compression along the armchair direction with an activation barrier of 203 meV/atom. Then the structure converts into the puckered structure through a bonding rearrangement process between the Sb atoms by overcoming an activation barrier of 145 meV/atom. The overall activation barrier of this two-step process is 241 meV/atom, which is expected to be accessible at the experimental temperatures. In addition, the kinetics of a nucleation-propagation process was studied by using a rectangular stripe supercell, as illustrated in Supporting Information Figure S9. A unit of the *α* phase first nucleates in a dH-Sb background by overcoming a barrier of 144 meV/atom, and propagates over the whole supercell subsequently through even smaller barriers of around 110 meV/atom. These results indicate the transition from dH to α phase is feasible and more likely to occur experimentally. Unfortunately, the intermediate state is not experimentally observed. According to our DFT calculation, the energy barriers required to overcome from the intermediate state to the α-Sb or dH-Sb are ~145 meV/atom, which is so small that the intermediate state can readily transform into the α- or dH state upon annealing at 350 K.

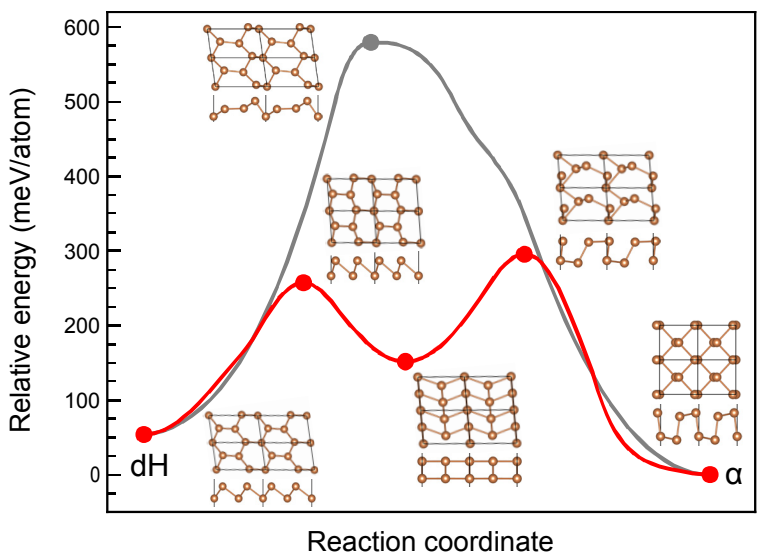

**Figure 4.** Kinetic pathways from the dH structure to the puckered α-Sb layer through a traditional direct process (grey path) and a two-step process (red path). Insets show the atomic structures of the initial dH structure, the intermediate structure, the final α-Sb structure, and the transition states. The points marked on the energy profile indicate the position of each corresponding atomic structure.

To exclude the impact of substrate on the formation of the hexagonal-like half-layer Sb, we explored a different substrate of $T_d$-MoTe$_2$ (Space group: $Pmn2_1$).[42] The STM images taken on Sb/MoTe$_2$ are shown in Supporting Information Figure S10, where the monolayer (1L) and bilayer (2L) α-antimonene, as well as the intermediate half-layer (1.5L) region, can be identified.

Consistent with the Sb on SnSe, the half-layer Sb formed on the $T_d$-MoTe$_2$ substrate also exhibits the hexagonal-like lattice symmetry. Moreover, such half-layer Sb can also form on top of the α-Sb monolayer instead of the MoTe$_2$ substrate. The similar phenomenon is also observed when directly depositing Sb over the α-Sb monolayer on SnSe (see Supporting Information Figure S11). Therefore, the influence of substrate to the half-layer formation can be ruled out.

A recent literature reported the substrate mediated phase transition from α-Sb to β-Sb monolayer on Bi$_2$Se$_3$ substrate.[43] In contrast, here we demonstrate the intrinsic growth kinetics of α-Sb without the influence of substrate.

**CONCLUSIONS:**

In summary, the growth mechanism of the α-antimonene monolayer has been investigated. Combining the experiment and DFT calculations, a two-step growth mode is unveiled. A distorted hexagonal atomic layer as the half-layer meta-structure is formed, prior to the transformation to the full puckered honeycomb monolayer. We believe such a two-step growth behavior should be universal to other monolayers in analogous structure, regardless of the choice of substrate.

**METHODS:**

The *α*-antimonene was grown and characterized in a commercial STM-MBE combined system (Unisoku USM-1500). The base pressure is $1\times10^{-10}$ Torr. The SnSe and $T_d$-MoTe$_2$ substrates were cleaved under ultrahigh vacuum, and the surface morphologies were verified *in situ* by STM prior to the sample growth. High purity (99.999%) Sb was loaded in a standard Knudsen evaporator. The flux was kept at ~0.3 monolayers (MLs) per minute. All the STM measurements were performed at 78 K with a mechanically polished PtIr tip. The sample was taken out of the STM stage and transferred to the preparation chamber prior to the post annealing process.

Density functional calculations were performed by using Vienna *ab Initio* simulation package (VASP).[44] The projector-augmented wave pseudopotentials with the Perdew-Burke-Ernzerhof type of generalized gradient approximation (GGA-PBE) were used.[45] The energy cut-off of the plane wave basis was set to 300 eV and a vacuum of more than 15 Å was introduced to the slabs. Van der Waals corrections of DFT-D2[46] method were treated in all vdW heterostructure systems. Optimized atomic structures were achieved until the forces on each atom were smaller than 0.01 eV/Å. The climbing-image nudged elastic band (cNEB)[47] method was used to simulate the phase transition process.

**AUTHOR CONTRIBUTIONS**

S.-C.L. conceived the project. Z.-Q.S., Q.-Q.Y., and C.-L.X. prepared the samples and carried out STM experiments with the assistance of Y.-J.X. and Z.-Y.J. Y.-Y.L. and Y.-B.C. grew the substrate single crystals. H.L. and W.Z. carried out theoretical calculations. Z.-Q.S., H.L., W.Z., and S.-C.L. prepared the manuscript. All authors discussed the results and commented on the manuscript.

**ACKNOWLEDGMENT**

This work was financially supported by the National Natural Science Foundation of China (Grants Nos. 11774149, 11790311, 11674299, and 11634011, 51872134, 11574131, 51902152), the Foundation for Innovative Research group of the National Natural Science Foundation of China (No. 51721001); the National Key Research and Development Program of China (Grant Nos. 2017YFA0204904 and 2019YFA0210004), the Strategic Priority Research Program of Chinese Academy of Sciences (Grant No. XDB30000000), Anhui Initiative in Quantum Information Technologies (Grant No. AHY170000), and the Fundamental Research Funds for the Central Universities (Grant Nos. WK2340000082 and WK2060190084).